\documentclass [a4paper, 11pt] {article}
\usepackage{amsmath}
\usepackage{amsfonts}
\usepackage{latexsym}
\usepackage{amsmath}
\usepackage{mathtools}
\usepackage{empheq}
\usepackage{graphicx}
\usepackage{caption}
\usepackage{subcaption}
\usepackage[toc,page]{appendix}
\usepackage{multicol}
\usepackage{color}
\begin{document}

\title{\textbf{Resonant absorption of kink oscillations in coronal flux tubes with continuous magnetic twist}}

\author{Zanyar Ebrahimi$^1$\thanks{E-mail: zebrahimi@maragheh.ac.ir} and
Karam Bahari$^2$\thanks{E-mail: karam.bahari@gmail.com}\\
$^1$\small{Research Institute for Astronomy $\&$ Astrophysics of
Maragha, University of Maragheh, Maragheh, Iran}\\
$^{2}$\small{Physics Department, Faculty of Science, Razi University, Kermanshah, Iran}}
\maketitle
\begin{abstract}
There are observational evidences for the existence of twisted magnetic field in the solar corona. Here, we have investigated resonant damping of the magnetohydrodynamic (MHD) kink waves in magnetic flux tubes. A realistic model of the tube with continuous magnetic twist and radially inhomogeneous density profile has been considered. We have obtained the dispersion relation of the kink wave using the solution to the linear MHD equations outside the density inhomogeneity and the appropriate connection formula to the solutions across the thin transitional boundary layer. The dependence of the oscillation frequency and damping rate of the waves on the twist parameter and longitudinal wavenumber has been investigated. For the flux tube parameters considered in this paper, we obtain rapid damping of the kink waves comparable to the observations. In order to justify this rapid damping, depending on the sign of the azimuthal kink mode number, $m=+1$ or $m=-1$, the background magnetic field must have left handed or right handed twisted profile, respectively. For the model considered here, the resonant absorption occurs only when the twist parameter is in a range specified by the density contrast.
\end{abstract}


\section{Introduction}
Uchida (1970) and Roberts et al. (1984) were among the first researchers who suggested the technique of coronal seismology to determine the equilibrium quantities of the solar corona. Uchida (1970) pointed out that on the basis of knowing the plasma density of the solar corona the magnetic field can be measured using the seismological diagnosis. The first model of the coronal flux tubes proposed by Edwin and Roberts (1983). They considered the magnetic flux tube as a density enhancement in a straight magnetic field and studied the dispersion diagram of the MHD waves for coronal and photospheric conditions and discussed the nature of the waves supported by the flux tube.

Until now various oscillation modes in the solar coronal loops have been observed and interpreted as the trapped MHD waves in magnetic flux tubes. Transverse oscillations in the coronal loops were first reported by Nakariakov et al. (1999) and Aschwanden et al. (1999) using the Transition Region and Coronal Explorer (TRACE) telescope on 1998 July 14 in the 171-{\AA} Fe IX emission lines. Nakariakov et al. (1999) interpreted these oscillations as MHD kink waves and indicated that the oscillations are strongly damped, such that the ratio of the damping time to the period of the oscillation is about 3-5.
Among the mechanisms believed to be responsible for the strong damping of the coronal loop oscillations, resonant absorption of the MHD waves is the most efficient mechanism. Resonant absorption of MHD waves established first by Ionson (1978), and is studied by some authors such as Hollweg and Yang (1988), Ruderman and Roberts (2002), Ofman (2009) and Morton and Erd\'{e}lyi (2009) to explain the strong damping of MHD kink waves. In this mechanism, the frequency of the global mode oscillation equals the background Alfv\'{e}n frequency at some radios inside the tube called resonance point which results to the cascading of the global mode energy to the local Alfv\'{e}n perturbations within a layer around the resonance point called resonance layer. For a good review on this topic, see e.g. Goossens et al. (2011).

Hollweg and Yang (1988) studied resonant absorption to explain damping of MHD waves in planar geometry. They were first to obtain approximate analytical expressions for the decay times of quasi-modes. They also applied their analytical results to kink modes $m=1$ in cylindrical geometry and established the fast damping of kink waves. Goossens et al. (1992) determined the conservation laws and the jump conditions across the slow and Alfv\'{e}n resonance points for a one dimensional cylindrical magnetic flux tube in the presence of plasma flow and magnetic twist. They used the jump conditions to study surface waves in solar magnetic flux tubes.
Goossens et al. (2002) mentioned that the quasi-mode kink oscillations in magnetic flux tubes can explain the observed rapid damping of the oscillations of coronal loops.
Goossens et al. (2009) investigated the nature of MHD kink waves in flux tubes. For this purpose they calculated the eigenfunctions, the frequency and the damping rate of MHD kink waves for three different MHD waves cases. They concluded that the kink waves do not care about the detailed qualities of the MHD wave environment, and if an adjective is to be used for the kink waves it should be Alfv\'{e}nic. They showed that in the resonance layer, pressure gradient force can be neglected only when the frequency of the kink wave does not differs much from the local Alfv\'{e}n frequency.
Giagkiozis et al. (2016) for the first time studied resonant absorption of axisymmetric MHD waves in twisted tubes. They implemented the conservation laws and derived a dispersion relation. They showed that the magnetic field and the density have a significant effect on the damping time of axisymmetric MHD waves.

As more properties of the waves in the solar corona becomes revealed, more structured models of the loops are needed to explain the physics of the waves. Verwichte et al. (2004) were first to detect the coexistence of the fundamental and first overtone of MHD kink waves in closed coronal loops and showed that the ratio of the period of the first overtone $P_2$ to the period of the fundamental mode $P_1$ is smaller than $2$. Andries et al. 2005 used longitudinal density stratification of the loop to explain the deviation of the period ratio from 2 (see also Erd\'{e}lyi and Verth 2007). This deviation can also be explained with a twisted magnetic field in the loop (see Erd\'{e}lyi and Carter 2006; Erd\'{e}lyi \& Fedun 2006 and Ebrahimi \& Karami 2016)

An interesting feature of the coronal flux tubes is that they may have a twisted magnetic field around the tube axis. Chae et al. (2000) suggested that a twisted magnetic field is needed in order to justify the rotational motions observed in coronal flux tubes. The magnetic twist of 14 coronal loops has been measured by Kwon\& Chae (2008). They found that the amount of the twist of the magnetic field, $\phi_{twist}$, was in the range $[0.22\pi, 1.73\pi]$. Here, $\phi_{twist}$ is the angel of rotation of the background magnetic field around the tube axis per characteristic length along the flux tube. It has been shown theoretically that if the amount of the magnetic twist in a flux tube exceeds a critical value which is about $\phi_c=2\pi$, the flux tube would be kink unstable (Shafranov 1957; Kruskal et al. 1958). For more details about the stability of kink waves in twisted flux tubes see e.g. Hood \& Priest (1979); Baty \& Heyvaerts (1996); Furno et al. (2006).

The effect of twisted magnetic field on the kink waves has been investigated for various background magnetic field and plasma conditions. In some models the azimuthal magnetic field varies discontinuously in the tube boundaries. Ruderman (2007) studied kink oscillations of a coronal loop in zero beta approximation in the existence of a twisted magnetic field in the internal region of the flux tube. For his model he showed that the magnetic twist does not affect the kink waves. Karami \& Bahari (2010) studied the effect of a twisted magnetic field on the resonant absorption of MHD waves by the assumption of incompressibility for the waves in a non-zero beta plasma. In order to have a twisted and continuous magnetic field in both the interior and exterior regions of the tube, for the sake of simplicity they considered an unphysical profile for the azimuthal component of the background magnetic field that goes to $\infty$ as $r\rightarrow\infty$.  Karami \& Bahari (2012) introduced an annulus to the model studied by Ruderman (2007) and showed that the magnetic twist of the annulus region can alter the oscillation frequency of the kink waves substantially. Based on this result they studied the ratio of the frequency of the first overtone and fundamental mode of kink wave. The azimuthal component of the background magnetic field considered by Karami \& Bahari (2012) was discontinuous at two different radiuses. Ebrahimi \& Karami (2016) investigated resonant absorption of MHD kink waves due to the existence of a twisted magnetic field in coronal flux tubes. They considered a discontinuous magnetic field in their work for the sake of simplicity. They showed that in the absence of a transition region for the plasma density in the radial direction the azimuthal component of the background magnetic field in flux tubes can introduce a strong damping for kink waves even in the limit of small twist values.
Bahari (2017) investigated the properties of standing MHD kink waves in a model coronal loop with an annulus in the presence of both magnetic twist and plasma flow. He showed that like plasma flow, the twisted magnetic field modifies the symmetry and phase difference of standing kink waves. He showed that depending on the direction of the magnetic twist and plasma flow, the presence of magnetic twist, like plasma flow, can cause underestimation or overestimation in obtaining the flow speed in coronal loops. Like Karami \& Bahari (2010) but in the zero beta approximation, Bahari \& Khalvandi (2017) considered the same unphysical profile for the azimuthal component of the background magnetic field in order to investigate the effect of twisted magnetic field on the nature of MHD kink waves.
Bahari (2018) has extended the work done by Soler et al. (2011) to the twisted loops. Since the kink waves in the long wavelength limit can be considered as incompressible waves (see Goossens et al. 2009), for simplicity Bahari (2018) assumed incompressible plasma. He discussed the symmetries in the phase diagram of the kink waves and showed that the damping length of the waves depends on the propagation direction, the amount of magnetic twist, the direction of magnetic twist and the azimuthal mode number of the waves.
In some models of coronal loops the azimuthal magnetic field varies continuously and is nonzero only in a region between internal and external regions of the loop.
Terradas \& Goossens (2012) calculated the eigenmodes of standing and propagating kink waves numerically for a coronal loop with twisted magnetic field. They concluded that in the presence of a twisted magnetic field, the polarisation of the transverse displacement of the plasma changes along the loop, and they suggested seismological application of this characteristic of the kink waves.
Ebrahimi et al. (2017) showed that depending on the direction of propagation of the wave and the twist profile considered in the tube, a twisted magnetic field can enhance or reduce the rate of phase-mixing of the perturbations of the MHD kink waves.
In some other models the magnetic twist is present in the whole loop and decreases away from the tube, which is physically more acceptable than the models with discontinuous magnetic field. Ruderman \& Terradas (2015) studied standing kink waves in the coronal loops with continuous equilibrium magnetic field. They showed that the effect of the magnetic twist on the period ratio of the kink waves depends on the density contrast of the tube.
Ruderman (2015) studied propagating MHD kink waves in twisted coronal loops and showed that in the presence of a twisted magnetic field the symmetry of the phase speed with respect to the propagation direction is broken. He called these waves accelerated and decelerated kink waves with phase speeds larger and smaller than the kink speed, respectively. He also stated that in the case of simultaneous identification of kink waves propagating with different phase speeds, this can be useful for coronal seismology because the amount of the magnetic twist can be estimated from the ratio of the frequencies.

To the authors it is interesting to study the properties of the MHD kink waves in a model flux tube which is more realistic than the models they studied before, a model with continuous magnetic twist and radially varying background density. For this purpose we add a thin inhomogeneous boundary layer to the model studied by Ruderman (2015) and investigate the oscillation frequency and damping rate of propagating components of standing kink waves. The paper is organized as follows: In section \ref{model} we introduce the model and the governing equations of MHD waves. In section \ref{solution}, we obtain the dispersion relation and in section \ref{results} we present the numerical results. Section \ref{Conclusions} is devoted to our conclusions.
\section{Model and Equations of Motion}\label{model}
Here, we approximate a real coronal flux tube by a straight cylinder with circular cross section of radius $R$. We use circular cylindrical coordinates ($r$,$\varphi$,$z$) such that the $z$ coordinate is aligned with the tube's axis. For simplicity we ignore the effect of gravity on the equilibrium density profile, hence we ignore the density stratification in the axial direction, also assume no equilibrium plasma flow in the flux tube. The profile of the equilibrium plasma density is assumed to be as follows
\begin{equation}\label{rho}
    \rho(r)=\left\{
    \begin{array}{ll}
        \rho_{\rm i},&r\leq a,\\
        \frac{1}{2}\left[\rho_{\rm i}+\rho_{\rm e}-\left(\rho_{\rm i}-\rho_{\rm e}\right)\sin\left(\frac{\pi}{2}\left(\frac{2r-R-a}{R-a}\right)\right)\right],&a<r<R,\\
        \rho_{\rm e},&r\geq R.
    \end{array}\right.
\end{equation}
Here $\rho_{\rm i}$ and $\rho_{\rm e}$ are the constant densities in the interior and exterior regions of the flux tube. The plasma density of the corona is about $10^{-15} gr.cm^{-3}$. The background magnetic field
is only a function of $r$ with $B_r=0$ and
\begin{equation}\label{bphi}
     B_{\varphi}(r)=\left\{
     \begin{array}{ll}
        Ar,& r\leq R,\\
        R^2 \frac{A}{r},& r>R.
     \end{array}\right.
\end{equation}
The constant $A$ is a parameter that controls the amount of twist in the flux tube. The profile of the azimuthal component of the background magnetic field is due to the existence of a constant current density inside the flux tube ($r<R$) along the tube axis. Here, we use the zero beta (ratio of the gas pressure to the magnetic pressure) approximation in our analysis. Hence, the background magnetic field is force free and satisfies the following equilibrium state condition
\begin{equation}\label{eqs}
     \frac{d B^2}{d r}=-2\frac{B_{\varphi}^2}{r}
\end{equation}
where $B^2=B_{\varphi}^2+B_z^2$. From Eq. (\ref{eqs}) and continuity of the magnetic pressure we obtain the $z$ component of the background magnetic field as
\begin{equation}\label{bz}
     B_{z}^2(r)=\left\{
     \begin{array}{ll}
        B_0^2+2A^2(R^2-r^2),& r\leq R,\\
        B_0^2,& r>R,
     \end{array}\right.
\end{equation}
where $B_0$ is a constant. The Magnetic field in the location of the coronal loops is of the order of $10 G$. It is clear from Eqs. (\ref{bphi}) and (\ref{bz}) that both the components of the background magnetic field are continuous at boundary of the flux tube. The azimuthal component of the magnetic field is zero at the tube axis and increases linearly in the internal part of the tube and in the external region of the tube decreases like $1/r$. The longitudinal component of the magnetic field takes its maximum value at the tube axis and decreases in the internal region of the tube. In the external region the longitudinal component of the magnetic field is constant. Since the background magnetic field is continuous everywhere, there is no surface current in our model. Models with a discontinuous magnetic field introduce a surface current at the location of discontinuity. The tearing mode instability can occur in a thin current sheet where the diffusive effects become important (e.g. Furth et al. 1963; Goldstone \& Rutherford 1995; Magara \& Shibata 1999; Ebrahimi \& Karami 2016). Even a small but non-zero resistivity makes the magnetic field lines tear and reconnect in the current sheet. In the present model this instability has been avoided by considering a continuous magnetic field.

The perturbations in the magnetic flux tube are governed by the linearized ideal MHD equations in zero beta approximation as follows
\begin{equation}\label{mhd1}
     \rho(r)\frac{\partial^2 \boldsymbol{\xi}}{\partial t^2}=\frac{1}{\mu_0}\{(\nabla\times{\mathbf B'})\times{\mathbf B} +(\nabla\times{\mathbf B})\times{\mathbf B'}\},
\end{equation}
\begin{equation}\label{mhd2}
    \mathbf{B'}=\nabla\times(\boldsymbol{\xi}\times\mathbf{B}),
\end{equation}
where $\boldsymbol{\xi}$ is the lagrangian displacement of the
plasma and $\mathbf{B'}$ is the Eulerian perturbation of the
magnetic field. Here, $\mu_0$ is the magnetic permeability
of the free space.

\begin{figure}
  \centering
    \includegraphics[width=70mm]{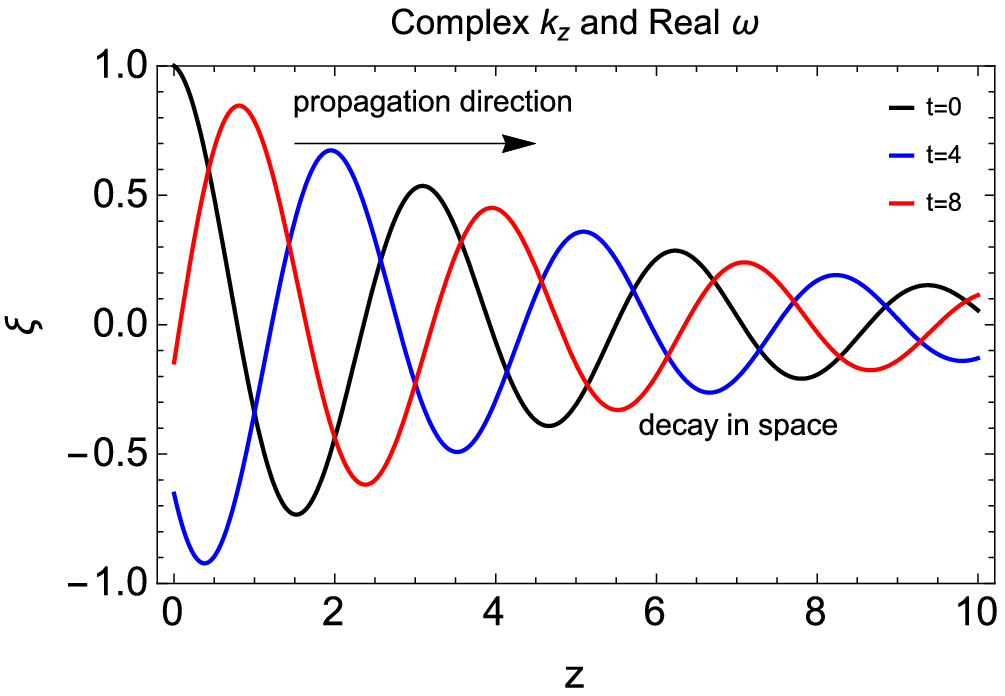}\\
    \includegraphics[width=70mm]{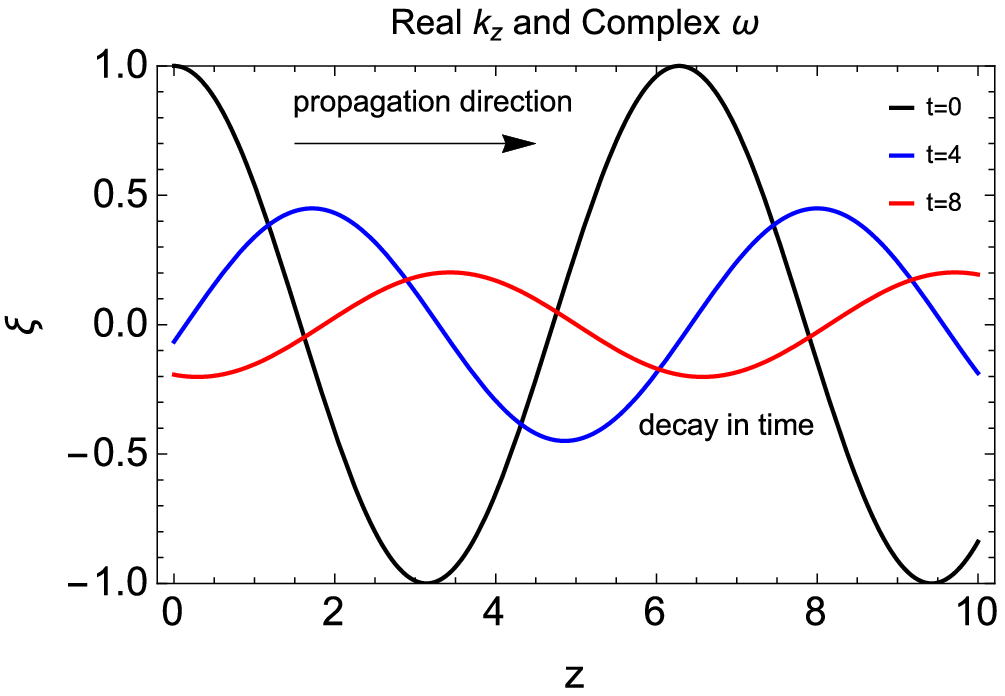}

\caption {top: A propagating wave that decays in space. bottom: A propagating wave that decays
in time.}
    \label{waves}
\end{figure}

A standing wave is a superposition of forward and backward propagating waves that have the same frequencies. As showed by Terradas and Goossens (2012), in the presence of a twisted magnetic field, the forward and backward propagating kink waves of the same frequency have different wave numbers. So it is not possible to Fourier analyze a standing kink wave in the longitudinal direction in the presence of a twisted magnetic field. In order to overcome this difficulty we study the individual forward and backward travelling components of a damped standing oscillation, separately.
Since the background quantities are not functions of $\varphi$ and $z$, we study the individual Fourier components of the perturbations and consider the $\varphi$ and $z$ dependence of the perturbations as $\exp \left[i\left(m\varphi+k_z z\right)\right]$, where $m$ and $k_z$ are azimuthal mode number and longitudinal wavenumber, respectively.
Here, we are looking for the global quasi mode solution of the waves that are damped in time due to resonant absorption and assume that the time dependence of the perturbations are of the form $\exp \left(-i\tilde\omega t\right)$ where the frequency $\tilde\omega$ is a complex quantity with dominant real part. Note that a propagating wave that starts from a source point at $z=0$ and decays in space as travels in the $z$ direction have a complex wavenumber and a real frequency. But the propagating components of a standing wave decay in time and have a real wavenumber and a complex frequency. Fig. \ref{waves} shows these two kinds of the damped propagation. Top panel of the figure exhibit a component of a propagating wave that decays in space at three different times. Bottom panel of the figure shows temporal damping of a travelling component of a standing wave. As illustrated in the figure, a component of a propagating wave with complex wavenumber and real frequency decays in the $z$ direction but a travelling component of a standing wave needs to have a real wavenumber and a complex frequency in order to decay in time.

Following Ruderman (2015), in thin tube approximation ($k_zR\ll1$) solutions of the Eqs. (\ref{mhd1}) and (\ref{mhd2}) for radial component of the Lagrangian displacement, $\xi_r$, and Eulerian perturbation of the total (magnetic) pressure, $P'$, in the internal ($r<a$) and external ($r>R$) regions of the flux tube are as follows

\begin{equation}\label{xir}
     \xi_r(r)=\left\{
     \begin{array}{ll}
        \xi_{r,\rm i}=\eta,& r\leq a,\\
        \xi_{r,\rm e}=\frac{\chi\sqrt{\mu_0\rho_{\rm e}}}{4R^2A\tilde\omega m}\ln\left|\frac{\left(\omega_{A, \rm e}-\tilde\omega\right)\left(B_0 k_z+\tilde\omega\sqrt{\mu_0\rho_{\rm e}}\right)}{\left(\omega_{A, \rm e}+\tilde\omega\right)\left(B_0 k_z-\tilde\omega\sqrt{\mu_0\rho_{\rm e}}\right)}\right|,& r>R,
     \end{array}\right.
\end{equation}
\begin{equation}\label{deltap}
     P'(r)=\left\{
     \begin{array}{lll}
        P'_{\rm i}&=r\eta\rho_{\rm i}\left(\tilde\omega^2-\frac{B_0^2 k_z^2-A^2}{\mu_0 \rho_{\rm i}}\right),& r\leq a,\\
        P'_{\rm e}&=\frac{\rho_{\rm e}\chi}{r}+\\
        &r\rho_{\rm e}\xi_{r,\rm e}\left[\tilde\omega^2-\frac{1}{\mu_0\rho_{\rm e}}\left(B_0^2 k_z^2-\frac{R^4 A^4}{r^4}\right)\right],& r>R,
     \end{array}\right.
\end{equation}
where
\begin{equation}\label{omegaAe}
     \omega_{A,{\rm e}}=\frac{1}{\sqrt{\mu_0\rho_{\rm e}}}\left(\frac{m A^2 R}{r^2}+k_z B_0\right).
\end{equation}
Here, $\eta$ and $\chi$ are constant coefficients that are determined by the appropriate boundary conditions. Notice that in the absence of the magnetic twist, Eqs. (\ref{xir}) and (\ref{deltap}) reduce to those obtained by Goossens et al. (2009) for pressureless flux tubes with uniform density in thin tube approximation.
\section{Connection Formula and Dispersion Relation }\label{solution}
In this section we aim to obtain the dispersion relation governing the MHD waves in the flux tube. For simplicity we assume thin boundary (TB) approximation which means that the thickness of the inhomogeneous boundary layer is much smaller than the radios of the tube. The TB approximation was first used by Hollweg and Yang (1988). Hence instead of solving the resistive equations of motion in the inhomogeneous boundary layer ($a<r<R$), the ideal solutions inside and outside the tube can be related to each other by the connection formula introduced by Sakurai et al. (1991). They showed that the jumps in the perturbations across the resonance layer are given by

\begin{eqnarray}
        \left[\xi_r\right]&=&-\frac{i\pi}{|\Delta|}\frac{g(r_A)}{\rho_{\rm i}B^2(r_A)}C_A(r_A),\label{jump1}\\
        \left[P'\right]&=&-\frac{i\pi}{|\Delta|}\frac{2B_\varphi(r_A)B_z(r_A)f(r_A)}{\mu_0\rho_{\rm i}r_A B^2(r_A)}C_A(r_A),\label{jump2}
\end{eqnarray}
where
\begin{eqnarray}
    \begin{aligned}\label{fgDelta}
    &C_A=g_B P'(r)-\frac{2f_B B_\varphi B_z}{\mu_0 r_A} \xi_r(r),\\
    &f=\frac{m}{r}B_\varphi+k_z B_z,\\
    &g=\frac{m}{r}B_z-k_z B_\varphi,\\
    &\Delta=\left.-\frac{\rm d}{{\rm d}r}\omega_A^2(r)\right|_{r=r_A},
    \end{aligned}
\end{eqnarray}
and $r_{A}$ is the location of the resonance point. Here
\begin{equation}\label{omegaA}
  \omega_A(r)=\frac{1}{\sqrt{\mu_0\rho(r)}}\left(\frac{m B_\varphi(r)}{r}+ k_z B_z(r)\right)
\end{equation}
 is defined as the background Alfv\'{e}n frequency.

Note that the expression for the jump in $P'$ (Eq. \ref{jump2}) has been derived for an equilibrium with continuously varying equilibrium quantities and the jump of $P'$ is due to the resonance in $r_A$ only. As explained by Goossens et al. (1992), in the absence of any resonance the radial component of displacement $\xi_r$ and the Lagrangian perturbation of total pressure
\begin{equation}\label{LP}
  \delta P=P'+\frac{dP_0}{dr}\xi_r=P'-\frac{B_\varphi^2}{r}\xi_r,
\end{equation}
have to be continuous. So, in the models with a discontinuous magnetic field at the tube boundary where $dP_0/dr$ is not continuous, an additional jump in $P'$ must be included in the right hand side of Eq. \ref{jump2}. As a particular case, in subsection 3.2 of Goossens et al. (1992) a model with both jumps at the boundary is considered. In the model studied by Bahari (2018) two boundaries exist in different radii, in one of them the resonance absorption takes place and $dP_0/dr$ is continuous hence only the jump due to resonance absorption is included, in the other one the resonance absorption does not take place but $dP_0/dr$ is discontinuous which again there is a jump in $P'$. In the present model in addition to the components of the magnetic field $B_\varphi$ and $B_z$ and the equilibrium magnetic pressure $P_0(r)$ also the derivative of the magnetic pressure, $dP_0/dr$, is continuous at the boundary, hence the jump is due to the Alfv\'{e}n resonance only and the connection formulae introduced by Sakurai et al. (1991) can be used to join the solutions inside and outside the tube.

Substituting the ideal solutions (\ref{xir}) and (\ref{deltap}) in the jump conditions
(\ref{jump1}) and (\ref{jump2}) and eliminating $\eta$ and $\chi$, we obtain the dispersion relation as
\begin{equation}\label{disp}
    d_0(\tilde\omega)+d_1(\tilde\omega)\left|_{r=r_A}\right.=0,
\end{equation}
where
\begin{eqnarray}\label{d0}
    \begin{aligned}
    d_0(\tilde\omega)=&-1+\frac{\sqrt{\mu_0\rho_{\rm e}}}{4Am\tilde\omega}\left[\left(\frac{a}{R}\frac{\rho_{\rm i}}{\rho_{\rm e}}-1\right)\tilde\omega^2+\right.\\
    &\left.\left(1-\frac{a}{R}\right)\left(\frac{B_0^2k_z^2-A^2}{\mu_0\rho_{\rm e}}\right)\right]\\
    &\times\ln\left|\frac{\left(\omega_{A, \rm e}-\tilde\omega\right)\left(B_0 k_z+\tilde\omega\sqrt{\mu_0\rho_{\rm e}}\right)}{\left(\omega_{A, \rm e}+\tilde\omega\right)\left(B_0 k_z-\tilde\omega\sqrt{\mu_0\rho_{\rm e}}\right)}\right|, \end{aligned}
\end{eqnarray}
and
\begin{eqnarray}\label{d1}
\footnotesize
    \begin{aligned}
     &d_1(\tilde\omega)=-\frac{i\pi}{|\Delta|}\frac{g}{\rho_{\rm i}B^2}\left[g r\rho_{\rm i}\left(\tilde\omega^2-\frac{B_0^2k_z^2-A^2}{\mu_0\rho_{\rm i}}\right)-\frac{2f B_\varphi B_z}{\mu_0 r}\right]\\
    & \times \left[\frac{\sqrt{\mu_0\rho_{\rm e}}}{4A m\tilde\omega}\left(\frac{2f B_\varphi B_z}{\mu_0 \rho_{\rm e} g r R}-\left(\tilde\omega^2-\frac{B_0^2k_z^2-A^2}{\mu_0\rho_{\rm e}}\right)\right)\right.\\
    &\left.\times\ln\left|\frac{\left(\omega_{A, \rm e}-\tilde\omega\right)\left(B_0 k_z+\tilde\omega\sqrt{\mu_0\rho_{\rm e}}\right)}{\left(\omega_{A, \rm e}+\tilde\omega\right)\left(B_0 k_z-\tilde\omega\sqrt{\mu_0\rho_{\rm e}}\right)}\right|-1\right].
    \end{aligned}
\normalsize
\end{eqnarray}
Here $\tilde\omega=\omega+i\gamma$, in which $\omega$ and
$\gamma$ are the oscillation frequency and the corresponding damping rate,
respectively. Note that in the limit of $a=R$ the resonance layer disappears, $\Delta\rightarrow\infty$ and the second term in the dispersion relation Eq. (\ref{disp}) disappears and Eq. (\ref{disp}) with real $\omega$ reduces to Eq. (48) of Ruderman (2015) for non-resonant MHD modes.

\section{Numerical results }\label{results}
In order to solve Eq. (\ref{disp}) we introduce the following dimensionless variables
\begin{equation}\label{dimles}
        \zeta\equiv\frac{\rho_{\rm i}}{\rho_{\rm e}},~~~\alpha\equiv\frac{AR}{B_0},~~~\epsilon\equiv k_zR.
\end{equation}
Also we rewrite all lengthes, magnetic fields and frequencies in new units $R$, $B_0$ and $\frac{B_0}{R\sqrt{\mu_0\rho_{\rm i}}}$, respectively. Hence, we obtain $d_0(\tilde\omega)$ and $d_1(\tilde\omega)$ in terms of the dimensionless variables as
\begin{eqnarray}\label{d0bar}
\small
    \begin{aligned}
    d_0(\tilde\omega)=&-1+\frac{\sqrt{\zeta}}{4\alpha m\tilde\omega}\left[\left(a-\frac{1}{\zeta}\right)\tilde\omega^2+\left(1-a\right)\left(\epsilon^2-\alpha^2\right)\right]\\
    &\times\ln\left|\frac{\left(\sqrt{\zeta}\left(m\alpha+\epsilon\right)-\tilde\omega\right)\left(\epsilon+\tilde\omega/\sqrt{\zeta}\right)}{\left(\sqrt{\zeta}\left(m\alpha+\epsilon\right)+\tilde\omega\right)\left(\epsilon-\tilde\omega/\sqrt{\zeta}\right)}\right|, \end{aligned}
\normalsize
\end{eqnarray}

\begin{eqnarray}\label{d1bar}
    \begin{aligned}
     &d_1(\tilde\omega)=-\frac{i\pi}{|\Delta|}\frac{g^2}{B^2}\left[\left(\tilde\omega^2-\epsilon^2+\alpha^2\right)r-\frac{2f B_\varphi B_z}{g r}\right]\\
    & \times \left[\frac{\sqrt{\zeta}}{4\alpha m\tilde\omega}\left(\frac{2f B_\varphi B_z}{g r}-\left(\frac{\tilde\omega^2}{\zeta}-\left(\epsilon^2-\alpha^2\right)\right)\right)\right.\\
    &\left.\times\ln\left|\frac{\left(\sqrt{\zeta}\left(m\alpha+\epsilon\right)-\tilde\omega\right)\left(\epsilon+\tilde\omega/\sqrt{\zeta}\right)}{\left(\sqrt{\zeta}\left(m\alpha+\epsilon\right)+\tilde\omega\right)\left(\epsilon-\tilde\omega/\sqrt{\zeta}\right)}\right|-1\right].
    \end{aligned}
\end{eqnarray}
Substituting Eqs. (\ref{bphi}) and (\ref{bz}) into Eq. (\ref{omegaA}) we obtain the dimensionless form of the background Alfv\'{e}n frequency as
\begin{equation}\label{omegaAdim}
    \begin{array}{ll}
     \omega_A(r)
    =\left\{
    \begin{array}{ll}
        \frac{1}{\sqrt{\rho(r)}}\left(m\alpha+\epsilon+ O(\epsilon\alpha^2)\right),&r\leq 1,\\
        \sqrt{\zeta}\left(\frac{m}{r^2}\alpha+\epsilon\right),&r> 1.
    \end{array}\right.
    \end{array}
\end{equation}
Note that here $\rho(r)$ is in unit of $\rho_{\rm i}$. In weak twist ($\alpha\ll1$) and thin tube ($\epsilon\ll1$) approximations, in Eq. (\ref{omegaAdim}) we can neglect terms of order $\epsilon\alpha^2$ respect to the other terms. Hence, the background Alfv\'{e}n frequency of the interior ($r<R$) of the flux tube becomes
\begin{equation}\label{omegaAidim}
  \omega_{\rm Ai}(r)\simeq\frac{1}{\sqrt{\rho(r)}}\left(m\alpha+\epsilon\right).
\end{equation}
Terradas and Goossens (2012) revealed that even if $B_\varphi\ll B_z$ i.e. $\alpha\ll 1$, the competition between $\alpha$ and $\epsilon$ determines the net effect of a twisted magnetic field on the MHD waves. They considered a piecewise step function profile for the plasma density in the radial direction and look for the non-resonant solutions in the presence of a continuous and twisted magnetic field. Since the profile of $B_\varphi$ chosen by them was parabolic it was possible to have resonance inside the flux tube even for a constant density profile. They restricted their calculations to small twist values in order to avoid resonance and kink instability in the flux tube. They also obtained the critical limit of the amount of twist above which the kink oscillation become resonant. Here the situations is different since there are a radial density stratification in a thin layer at the tube boundary that introduces the resonance for the kink waves even for zero twist. Since $B_\varphi$ has a linear profile and $B_z$ is almost constant (to first order in $\alpha$) inside the tube, the occurrence of resonance is not related to the magnetic field. However, in what follows the results show that the existence of a twisted magnetic field may suppress the resonance caused by the radial density stratification for magnetic twist values higher than a critical limit.

It is clear from Eq. (\ref{omegaAidim}) that for $m=+1$ and $\alpha=-\epsilon$ or $m=-1$ and $\alpha=+\epsilon$, the background Alfv\'{e}n frequency inside the tube is $\omega_{\rm Ai}=0$. For example, if we  insert $m=1$ and $\alpha=-\epsilon$ in the non-resonant dispersion relation $d_0=0$ obtained by Ruderman (2015), the obtained frequency is $\omega=0$. The same result is obtained for $m=-1$ and $\alpha=\epsilon$. So, in these cases the kink waves are not allowed to propagate in the flux tube. In fact, in these cases the wave vector which is defined as ${\mathbf k}\equiv\frac{m}{r}\hat{\varphi}+k_z\hat{z}$ is perpendicular to the background magnetic field i.e. $\mathbf{k}\cdot \mathbf{B}=0$. For an incompressible plasma $\boldsymbol{\nabla}\cdot\boldsymbol{\xi}=0$ when $\mathbf{k}\cdot \mathbf{B}=0$, the right hand side of Eq. (\ref{mhd2}) vanishes and the resistive effects become important. In this case, the resistive diffusion results to the so called resistive kink instability, that is a reconnecting process (e.g. Biskamp 2000; Wesson 2004; Priest 2014).

Here we take the parameters of a typical oscillating coronal flux tube by the following considerations:
\begin{itemize}
    \item As stated by Aschwanden et al. (2003), the typical values of the density ratio $\zeta$ for coronal flux tubes are in the range $[2,10]$. Here, we take the extreme values $\zeta=2$ and $\zeta=10$ in our calculations.
    \item We investigate both forward ($k_z>0$) and backward ($k_z<0$) propagating components of a standing MHD kink wave for both the right-hand ($\alpha>0$) and the left-hand ($\alpha<0$) twisted magnetic fields.
    \item Since we used the thin tube approximation ($\epsilon=k_zR\ll1$), the results are obtained for $\epsilon$ in the range $[-\frac{\pi}{50},\frac{\pi}{50}]$.
    \item If the magnetic twist value in the loop, which here is defined as
    \begin{equation}\label{phitwist1}
    \phi\equiv\left.\frac{L}{R}\frac{B_\varphi}{B_z}\right|_{r=R}= 2\pi N_{twist}
    \end{equation}
     exceeds a critical value $\phi_c$, then the loop becomes kink unstable (e.g. Shafranov 1957; Kruskal et al. 1958). Here, $L$ is the length scale of the wave propagation in the $z$ direction which we assume to be equal to the wavelength ($\lambda$) and $N_{twist}$ is the number of twist turns per length $L$. Substituting $L=2\pi/k_z=\lambda$ in Eq. (\ref{phitwist1}) we get
     \begin{equation}\label{phitwist2}
        \phi=2\pi\frac{\alpha}{\epsilon}.
     \end{equation}
     So, there is a constraint on the maximum value of the twist parameter $\alpha$ in order to the kink oscillations of the flux tube be stable. Note that Ruderman (2015) derived solutions (\ref{xir}) and (\ref{deltap}) in the weak twist approximations. He assumed $\alpha\leq\epsilon$ (i.e. $\phi\leq\phi_c= 2\pi$) in order to avoid the kink instability in the flux tube. Hence, the dispersion relation obtained in this paper is valid in this limit. Terradas and Goossens (2012) obtained the critical value of the magnetic twist from equalizing the contribution of the azimuthal and longitudinal components of the magnetic field in the background Alfv\'{e}n continuum. They found that in order to have a stable kink oscillation the amount of twist must be smaller than $\pi$. Since they considered the longitudinal length scale as $L=\pi/k_z=\lambda/2$, the critical twist obtained by them is equivalent to $\phi_c=2\pi$ that we use in this paper.

     With these considerations, for a given $\epsilon\in[-\pi/50,\pi/50]$, from Eq. (\ref{phitwist2}) we deduce that in order to have valid results, the twist parameter must be in the range
     \begin{equation}\label{alphamax}
        -\epsilon\leq\alpha\leq\epsilon.
     \end{equation}
    From Eq. (\ref{phitwist2}) we have
    \begin{equation}\label{3}
      \epsilon=\frac{2\pi}{\phi}\alpha.
    \end{equation}
    This equation shows that any linear relation between $\epsilon$ and $\alpha$ corresponds to a constant $\phi$. In other words, if we adopt a constant $\phi$ in the flux tube, there exists a set of infinite pairs of $\epsilon$ and $\alpha$ satisfying Eq. (\ref{3}) that make a straight line in the $\alpha-\epsilon$ plane. For example, for $m=+1$ if $\phi=2\pi$ (right handed twist) then we have $\alpha=\epsilon$ line (dashed lines in the figures) and if $\phi=-2\pi$ (left handed twist) it yields $\alpha=-\epsilon$ line (dashed-dotted line in the figures). These lines are the borders between the kink stable and kink unstable regions. In the contour plots a point with $|\alpha|>|\epsilon|$ that is equivalent to $|\phi|>2\pi$ is in the kink unstable region.\\
    \item In thin boundary approximation we can estimate the location of the resonance as $r_A\simeq(a+R)/2$.
\end{itemize}

\begin{figure}
  \centering
    \includegraphics[width=80mm]{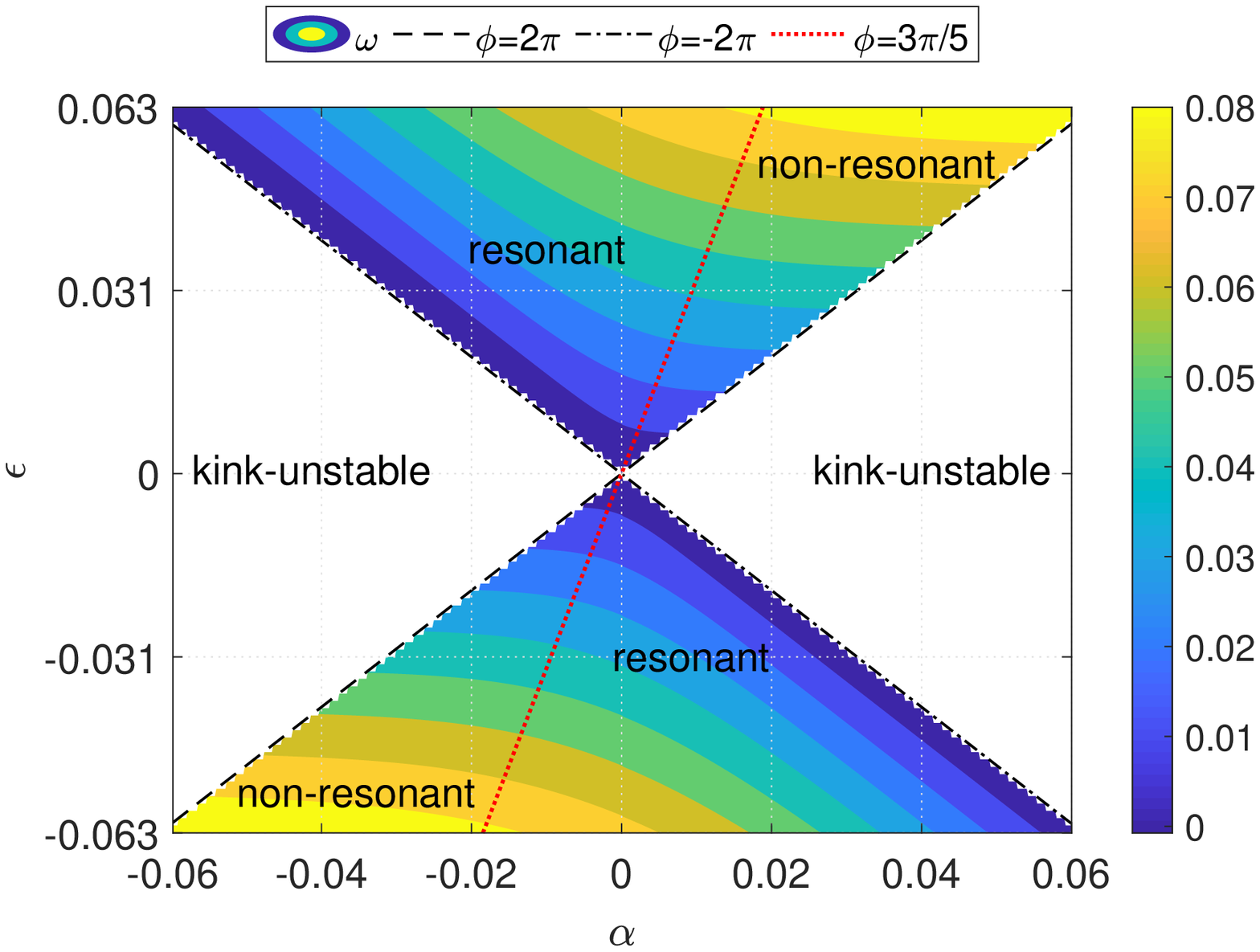}\\
    \includegraphics[width=80mm]{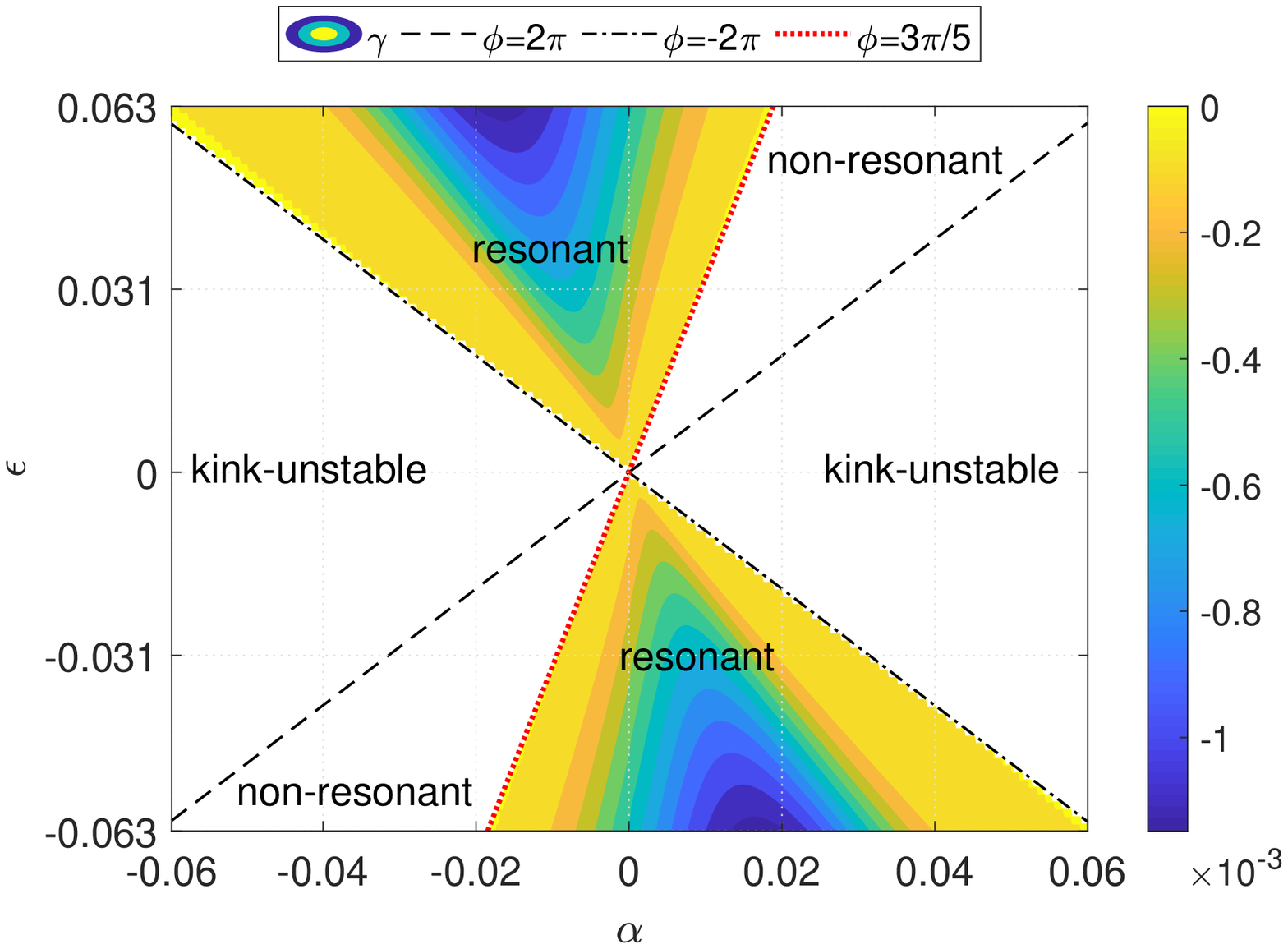}\\
    \includegraphics[width=80mm]{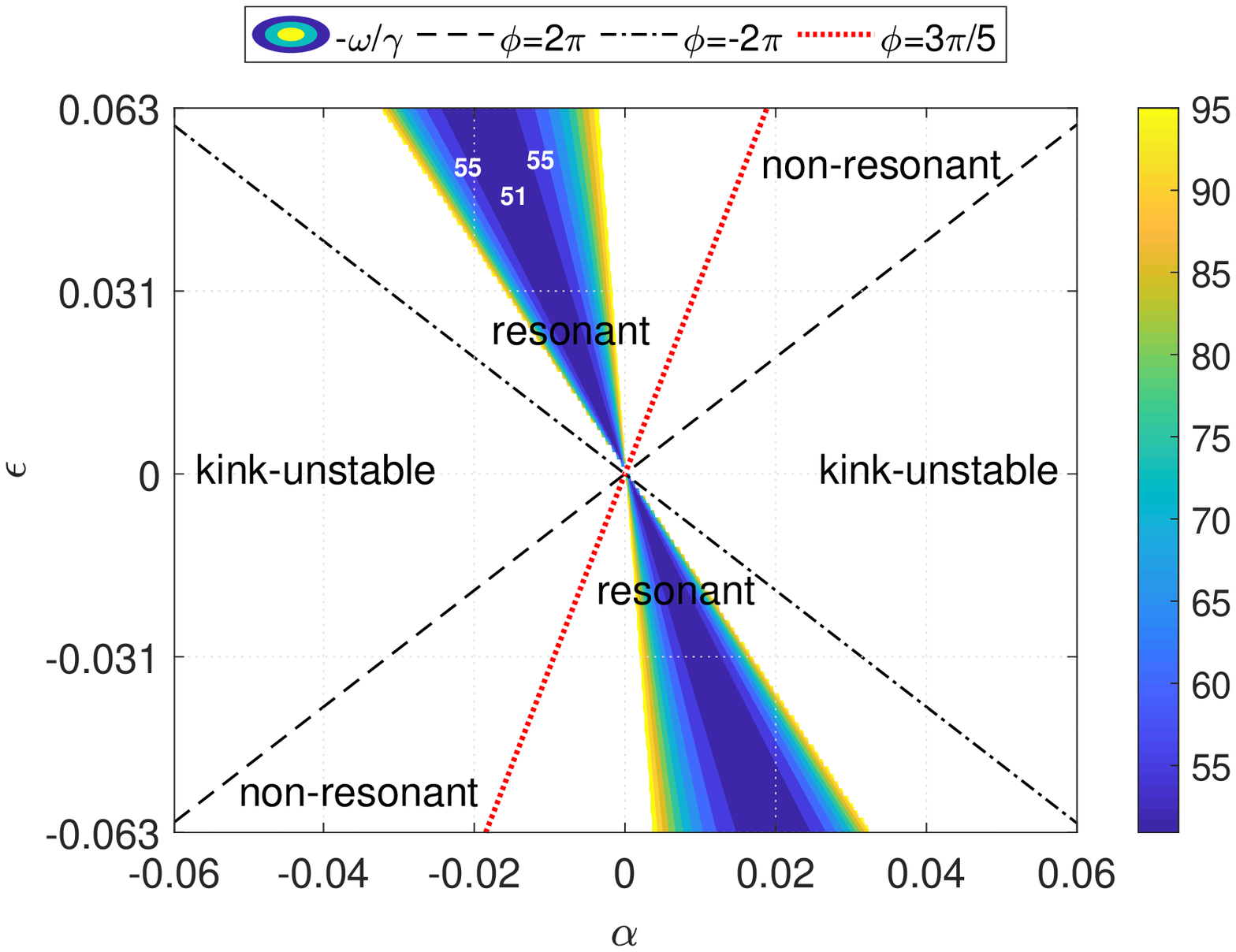}

\caption {Contour plots of the frequency (top), the damping rate (middle) and the ratio of the frequency to the damping rate (bottom) versus $\alpha$ and $\epsilon$ for $m=1$ and $\zeta=2$. The dashed and dashed-dotted lines illustrate $\phi=2\pi$ and $\phi=-2\pi$, respectively. The red dotted line separate the resonant and non-resonant regions. }
    \label{omegagamma2Dz2}
\end{figure}
\begin{figure}
  \centering
    \includegraphics[width=80mm]{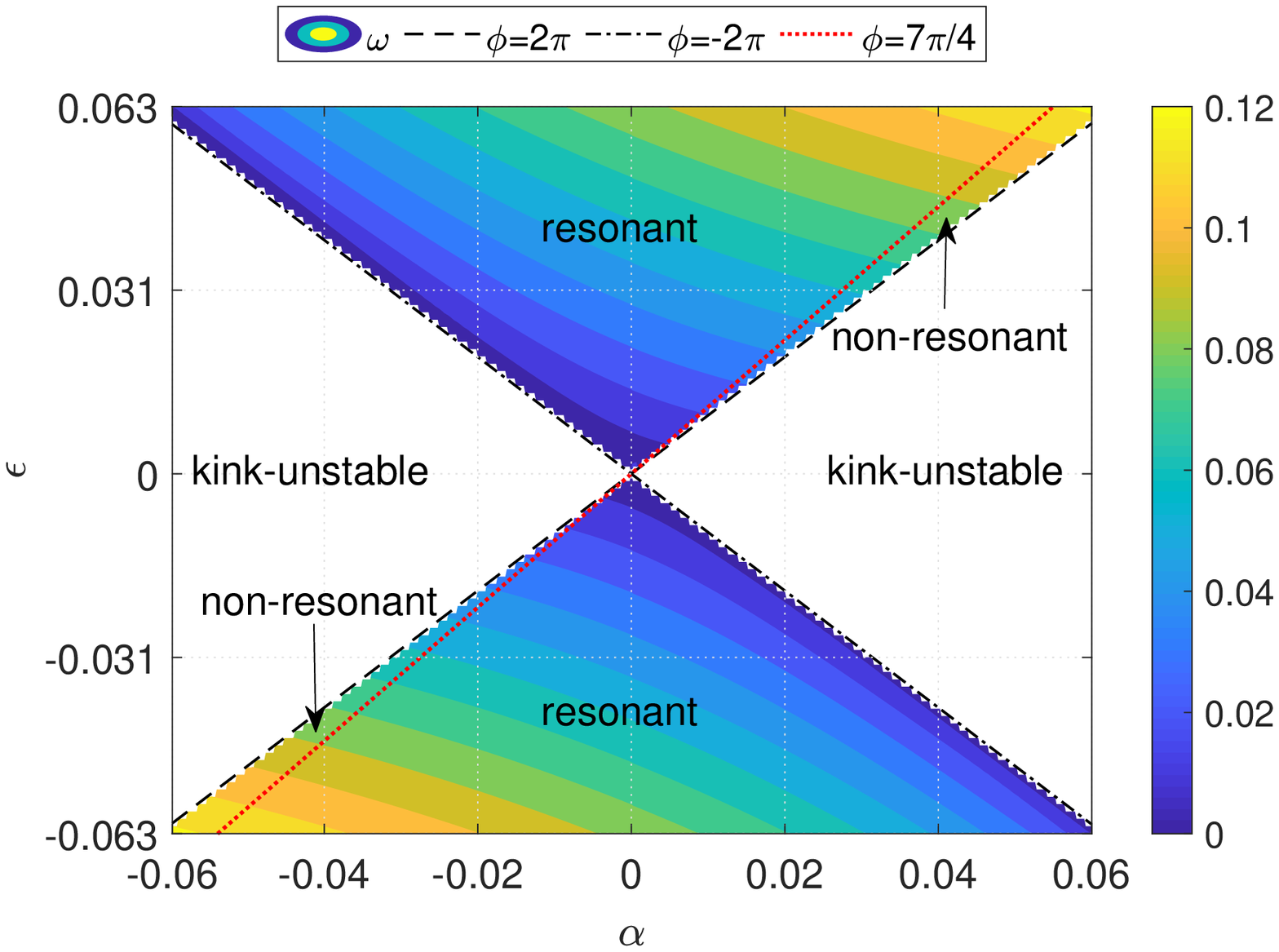}\\
    \includegraphics[width=80mm]{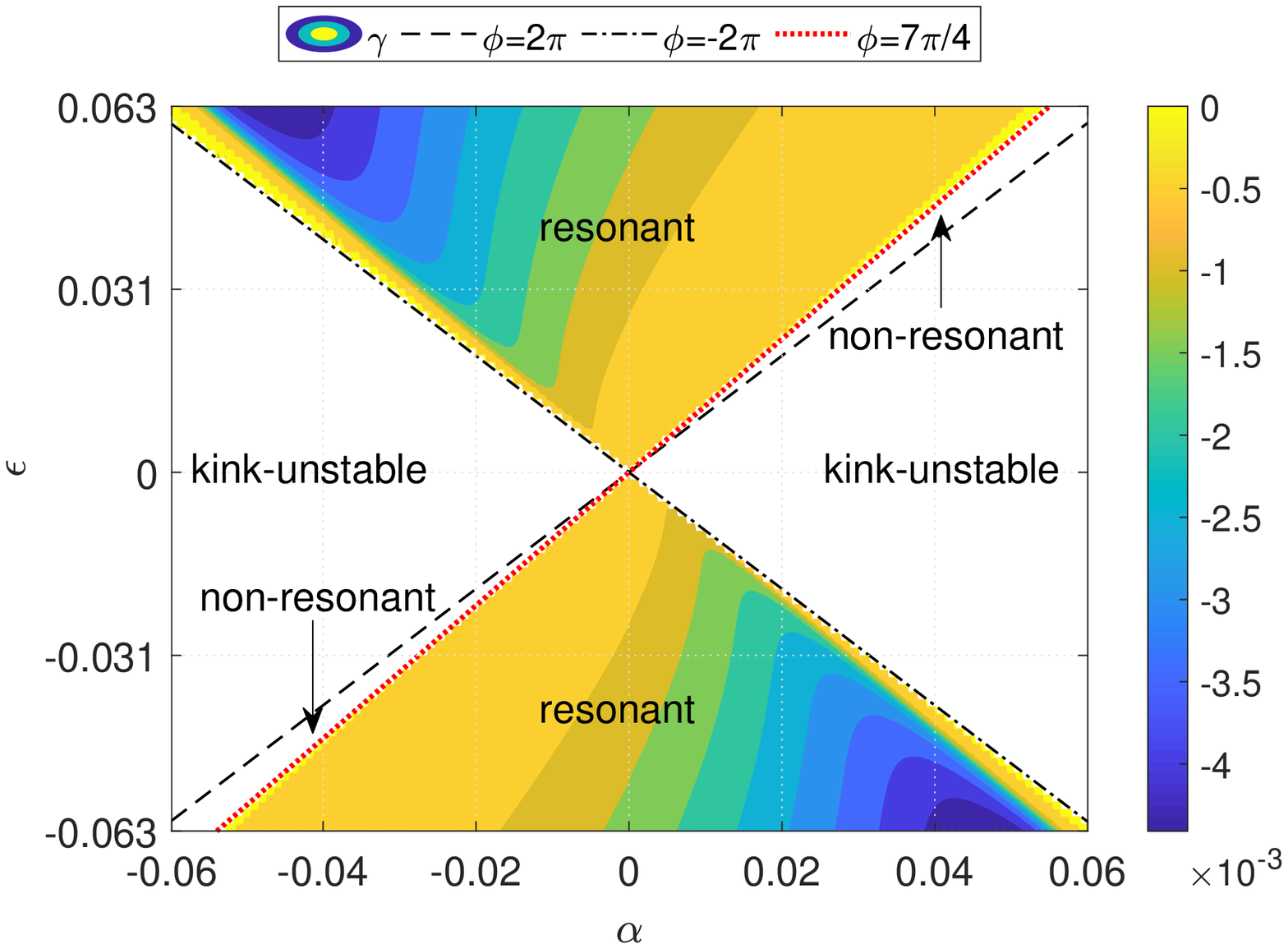}\\
    \includegraphics[width=80mm]{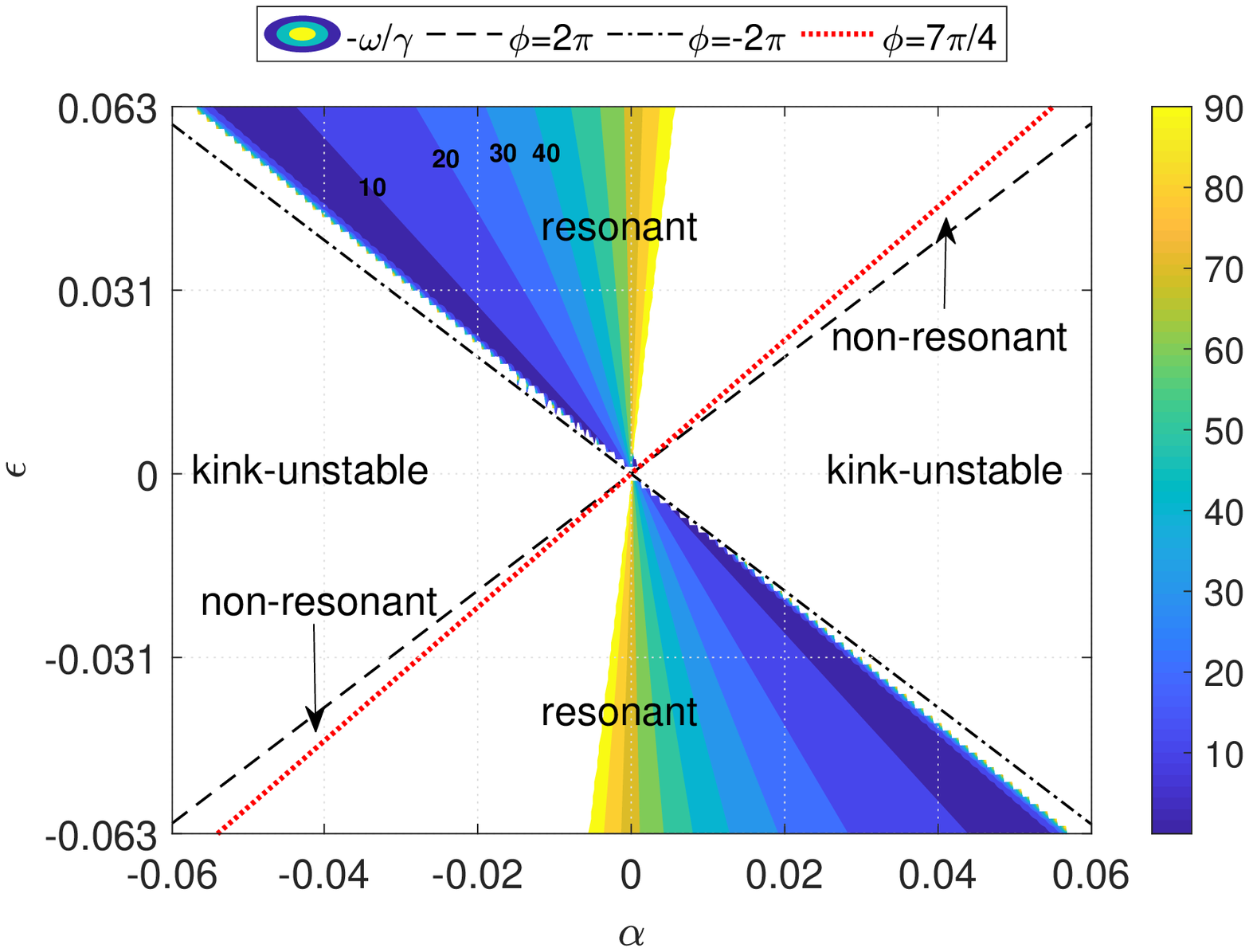}

\caption {Same as Figure \ref{omegagamma2Dz10} but for $\zeta=10$.}
    \label{omegagamma2Dz10}
\end{figure}

With these considerations we solve Eq. (\ref{disp}) and find its complex roots numerically in the kink stable regions. We have obtained the oscillation frequency $\omega$ and the damping rate $\gamma$ of the kink waves as a function of $\alpha$ and $\epsilon$ for $m=+1$. Figs. \ref{omegagamma2Dz2} and \ref{omegagamma2Dz10} show the contour plots of the oscillation frequency, the damping rate and the ratio of the frequency to the damping rate in the $\alpha-\epsilon$ plane. Fig. \ref{omegagamma2Dz2} is for $\zeta=2$ and Fig. \ref{omegagamma2Dz10} is for  $\zeta=10$. In both figures the dashed and dashed-dotted lines in the plots correspond to $\phi=2\pi$, and $\phi=-2\pi$ respectively. As illustrated in the figures, we do not plot the results in the kink-unstable regions where the magnitude of the magnetic twist, $|\phi|$, is larger than its critical value $\phi_c=2\pi$. Hence, we have plotted the results in the regions which are valid according the stability criterion. It is clear from the top panel of Figs. \ref{omegagamma2Dz2} and \ref{omegagamma2Dz10} that in the both cases of $\zeta=2$  and $\zeta=10$ when $\epsilon>0$ ($\epsilon<0$) with increasing (decreasing) the twist parameter, $\alpha$, the frequency of the kink wave increases. This results are in agreement with the results shown in the top panel of figure 1 in Bahari (2018).
Also it is clear from the top panel of Figs. \ref{omegagamma2Dz2} and \ref{omegagamma2Dz10} that in the both cases of $\zeta=2$  and $\zeta=10$, for a given value of $\epsilon$, the frequency of the kink waves is a monotonic function of $\alpha$ that starts from $\omega=\omega_{\rm Ai}=0$ at a point on the line $\phi=-2\pi$ (i.e $\epsilon=-\alpha$) (the dashed-dotted line in the figures). This means that for the model considered in this paper, when $m=+1$ and $\phi=-2\pi$ the kink wave do not propagate in the flux tube.

Our results show that if the amount of the magnetic twist $\phi$ exceeds a critical value $\phi^*$, the real part of the obtained frequency falls below the background Alfv\'{e}n frequency continuum (i.e $\omega<\omega_{Ai}$). In this case the oscillation is non-resonant. Note that the critical point that distinct the resonant and non-resonant regions, is determined by $\phi$ not $\epsilon$ neither $\alpha$ alone. The obtained value of $\phi^*$ for $\zeta=2$ and $\zeta=10$ is $\phi^*\simeq3\pi/5$ and $\phi^*\simeq7\pi/4$, respectively. As mentioned before, from Eq. (\ref{3}) we see that each of these constant values of $\phi^*$ corresponds to a straight line in the $\alpha-\epsilon$ plane. The red dotted line in Figs. \ref{omegagamma2Dz2} and \ref{omegagamma2Dz10} correspond to $\phi=3\pi/5$ and $\phi=7\pi/4$ respectively, that exhibit the location where $\omega=\omega_{\rm Ai}$. This result is interesting because it shows that if the twist parameter $\phi$ becomes larger than a certain value specified by the density contrast, there exists an undamped kink wave in the presence of inhomogeneous boundary layer, and resonant absorption cannot be considered as the damping mechanism for this wave. Note that since our model is different form the model studied by Terradas and Goossens (2012) there is differences between the conditions in which resonant absorption takes place. In Terradas and Goossens (2012) the twist has been assumed to be small enough to the waves be undamped, but in our model in order to the resonant absorption does not take place and the waves be undamped, the twist must be assumed to be large enough.

The damping rate of the kink waves in the tubes with density contrasts $\zeta=2$ and $\zeta=10$ are shown in the middle panels of Figs. \ref{omegagamma2Dz2} and \ref{omegagamma2Dz10} respectively. As illustrated in the figures, for $\zeta=2$ and $\zeta=10$ the damping rate is plotted in the ranges $(-2\pi,3\pi/5)$ and $(-2\pi,7\pi/4)$, respectively. It is clear that for any value of $\epsilon$ the damping rate is not a monotonic function of $\alpha$. As the middle panels of Figs. \ref{omegagamma2Dz2} and \ref{omegagamma2Dz10} show, for both cases of $\zeta=2$ and $\zeta=10$, the damping rate of the resonant kink wave near the $\phi=-2\pi$ line is $\gamma=0$, and with increasing $\phi$ from $-2\pi$ the damping rate increases and reaches a maximum value and then decreases again to $\gamma=0$ at $\phi=3\pi/5$ and  $7\pi/4$, respectively.

The absolute value of the ratio of the frequency to the damping rate has been illustrated in the bottom panels of Figs. \ref{omegagamma2Dz2} and \ref{omegagamma2Dz10}. Note that we only show the results for $-\omega/\gamma < 100$. In order to justify the rapid damping of MHD kink waves ($\omega/(2\pi\gamma)=\tau_D/P\simeq3-5$) reported by the observations, the ratio $-\omega/\gamma$ must be in the range $(20,30)$. As Fig. \ref{omegagamma2Dz10} shows, in the case of $m=+1$ and $\zeta=10$, this happens in a narrow triangular shape region where $\alpha$ and $\epsilon$ have opposite signs. As Fig. \ref{omegagamma2Dz2} shows, the minimum value of $-\omega/\gamma=51$ that is equivalent to $\tau_D/P\simeq 8$ is comparable with observational values. So, the model considered in this paper can justify the rapid damping of the MHD kink waves for specific pairs of $\alpha$ and $\epsilon$ that have opposite signs if $m=+1$. The bottom panels of Figs. \ref{omegagamma2Dz2} and \ref{omegagamma2Dz10} also reveal that the contour lines of the plots of $\omega/\gamma$ coincide with $\phi=const$ lines. Previous results obtained by e.g. Ruderman \& Roberts (2002) show that in the absence of the magnetic twist, the ratio $\omega/\gamma$ is not a function of $\epsilon$ that is a special case of the results obtained here for $\phi=0$. Our results show that in general, the ratio $\omega/\gamma$ remain unchanged when the twist value is constant in the flux tube. In other words, for a given $\alpha\neq0$ the ratio $\omega/\gamma$ is a function of $\epsilon$, but if $\alpha$ and $\epsilon$ vary together so that the twist value remain unchanged, it does not affect the ratio $\omega/\gamma$. This is an important results which should be considered in the seismology of coronal flux tubes, since for a given $\zeta$ by measuring the ratio of the damping time to the period of a decaying kink wave, we can estimate the amount of the magnetic twist $\phi$ in the tube.

Terradas and Goossens (2012) showed that it is possible to construct a standing kink wave in the presence of a twisted magnetic field from two forward and backward propagating kink waves of different longitudinal wavenumbers that have the same frequencies and propagate in opposite directions. They studied undamped kink waves and both of the wavenumber and frequency of the waves were real. But here the damped kink waves have complex frequencies. Our results show that even if we can find two propagating waves that have equal frequencies their damping rate are not the same and as a results it is not possible to make a standing kink oscillation from two propagating damping kink waves in the presence of a twisted magnetic field. This result is an interesting subject to study that is beyond the goals of this paper.

\section{Conclusions}\label{Conclusions}
Here, we investigated damping of standing MHD kink waves in a coronal flux tube in the presence of a twisted and continuous magnetic field. To this aim we obtained a dispersion relation for the individual propagating components of the standing kink wave. In the considered model resonant absorption occurs as a result of the existence of a thin layer of density inhomogeneity at the surface of the tube that connects smoothly the constant densities of the interior and exterior regions. The kink stability of the flux tube depends on both of the twist parameter and the longitudinal wave number. So, in order to investigate the effect of the twisted magnetic field on the MHD kink waves, we used the contour plots of the results versus parameters $\alpha$ and $\epsilon$. The results show that:
\begin{itemize}
    \item For a given $\epsilon$, the frequency of the kink waves is a monotonic function of the twist parameter even when the range of the variation of the twist parameter include both positive and negative values.
        But for a given value of $\epsilon$ the damping rate is not a monotonic function of the twist parameter.
    \item Resonant absorption occurs only when the value of the magnetic twist $\phi$ is in a range specified by the density contrast of the tube.
    \item The magnitude of the ratio of the frequency to the damping rate ($\omega/\gamma$) is not a monotonic function of $\alpha$ or $\epsilon$ in general. The results show that for the parameters considered in this study ($m=1$, $\zeta=2, 10$), $\alpha$ and $\epsilon$ must have opposite signs in order to justify the rapid damping of the kink oscillations in coronal flux tubes.
    \item For $\alpha=0$, the ratio $\omega/\gamma$ is not affected by $\epsilon$ that is in good agreement with the previous results obtained by e.g. Ruderman \& Roberts (2002). Our results show that in general, $\omega/\gamma$ is a function of $\phi$ and remain unchanged when $\phi=const$ for different values of $\epsilon$ and $\alpha$.
    \item Our results show that for the model considered in this paper,
        in contrast to the undamped kink waves (see Terradas and Goossens 2012), it is not possible to construct a resonantly damped standing kink wave from two forward and backward propagating kink waves with the same frequencies, since the corresponding damping rates are always different.

\end{itemize}





\end{document}